\documentclass[10pt,journal,final]{IEEEtran}
\usepackage[T1]{fontenc}
\usepackage{amsthm,amsmath,amssymb,mathtools,bm}
\usepackage{stfloats}
\usepackage{graphicx}
\usepackage{subfigure}
\usepackage{cite}
\usepackage[ruled,linesnumbered]{algorithm2e}

\usepackage{algpseudocode}
\makeatletter
\newcommand{\removelatexerror}{\let\@latex@error\@gobble}
\makeatother

\ifCLASSINFOpdf
\else
\fi

\begin{document}
	
	\title{A Low Complexity MAP Detector for OTFS Modulation in Logarithmic Domain}
	
	\author{Haoyan~Liu,
		Yanming~Liu,
		and Min~Yang
		\thanks{The authors are the School of Aerospace Science and Technology, Xidian University, Xi’an 710071, China.}
	}
	
	\maketitle
	
	\begin{abstract}
		Orthogonal time-frequency space (OTFS) has been confirmed to take advantage of full time-frequency diversity to significantly improve error performance in high-mobility scenarios. We found that the proposed message passing (MP) and variational Bayes (VB) detectors can achieve approximate maximum a posteriori (MAP) detection, the interferences cannot be completely eliminate in the absence of noise. To achieve near-optimal MAP detection, this letter proposes a novel detection method based on sum-product algorithm (SPA) with low complexity. Leveraging subtly factorized posteriori probabilities, the obtained pairwise interactions can effectively avoid enumeration of high-dimensional variables, thereby making it applicable to fractional Doppler cases. We further simplify the proposed algorithm in the logarithmic domain so that the message propagation processing only involves addition. Finally, simulations results demonstrate the superior error performance gains of our proposed algorithm at high signal-to-noise ratios (SNRs).
		
	\end{abstract}
	
	\begin{IEEEkeywords}
		Orthogonal time frequency space (OTFS), low complexity, sum-product algorithm (SPA), logarithmic domain.
		
	\end{IEEEkeywords}

	\IEEEpeerreviewmaketitle
	
	\section{Introduction}
	\IEEEPARstart{F}{uture} cellular communications are envisioned to support reliable transmission in high-mobility scenarios, such as high speed trains and unmanned aerial vehicles \cite{wong_schober_ng_wang_2017}. As a multiplexing scheme with high spectral efficiency, orthogonal frequency division multiplexing (OFDM) can mitigate the effect of inter-symbol interferences (ISI) in time-invariant frequency selective channels. However, Doppler shift will destroy the orthogonality of subcarriers and lead to inter-carrier interferences (ICI), which significantly degrades the performance of OFDM.
	
	Orthogonal time frequency space (OTFS) modulation is a recently proposed scheme to combat Doppler shifts in multipath wireless channels \cite{7925924}. It can be equivalently considered as the pre-processing technology of OFDM, in which information symbols are modulated in delay-Doppler domain, and then spread in time-frequency domain using Heisenberg transform. It can be shown that all symbols over a transmission frame experience the identical channel response in delay-Doppler domain. Consequently, OTFS can take advantage of the potential channel diversity to have superior error performance compared to OFDM in high Doppler environments \cite{8892482}.
	
	To achieve full diversity gain, the optimal maximum a posteriori (MAP) detector is required at the receiving end. At present, one of the most popular approximate MAP detector is the message passing (MP) algorithm \cite{8424569}. By approximating the interferences with the Gaussian assumption, the MP detector achieves a linear complexity with the number of symbols. An alternative variational Bayes (VB) detector was proposed in \cite{9082873}. The VB detector does not need to evaluate the covariance matrix, thus resulting in a lower complexity than that of the MP detector. However, we found a common problem that interferences cannot be completely eliminated due to the independence assumption adopted by both MP and VB detectors, and error floor will occur at high signal-to-noise ratios (SNRs). 
	
	In this letter, we design a low complexity MAP detector based on the framework of sum-product algorithm (SPA) \cite{910572}. As an exact inference approach, SPA can effectively prohibit the error floor phenomenon. Nevertheless, it is known that the enumeration yields exponential complexity with respect to the number of connections over factor graph. A similar work was proposed in \cite{li2020hybrid}, the authors developed a novel hybrid MAP detection method to reduce the SPA complexity, but it assumes integer Dopplers and still has exponential complexity. As for our scheme, it only requires to enumerate one variable by using subtly factorized posteriori probabilities and achieves a linear complexity, thereby providing a feasible approach for the case of fractional Doppler. Another advantage is that there only involves addition in message propagation by utilizing some mathematic tricks. Simulation results show that our proposed method dramatically outperforms VB detector at high SNRs and will only has slight performance loss at low SNRs.

	\section{System Model}
	
	In this section, we review the basic OTFS systems with one transmit and one receive antenna. A sequence of information bits is mapped to $N \times M$ data symbols $x\left[k,l\right]$ in the delay-Doppler domain with constellation set $\mathcal{A}$, where $k=0,1, \cdots, N-1$, $l=0,1, \cdots, M-1$ denote the Doppler and delay indices, respectively. The OTFS converts $x\left[k,l\right]$ to symbols $X\left[n,m\right]$ in the time–frequency domain using inverse symplectic finite Fourier transform (ISFFT), given by
	\begin{equation}
	X\left[n,m\right] = \frac{1}{\sqrt{NM}}\sum^{N-1}_{n=0}\sum^{M-1}_{m=0} x\left[k,l\right]e^{j2\pi \left(\frac{nk}{N}-\frac{ml}{M}\right)}.
	\end{equation}
	The obtained $X\left[n,m\right]$ are further modulated on a set of bi-orthogonal time-frequency basis functions for multiplex transmission, 
	\begin{equation}
	s(t) = \sum_{n=0}^{N-1} \sum_{m=0}^{M-1}X\left[n,m\right]g_{tx}(t-nT)e^{j2\pi m\Delta f(t-nT)},
	\end{equation}
	The above equation is also called Heisenberg transformation, where $g_{tx}(t)$, $T$ and $\Delta f$ denotes the normalized prototype pulse, symbol period and subcarrier separation, respectively. Suppose there are $P$ independent scattering  paths in signal propagation, the delay-Doppler channel representation is given by
	\begin{equation}
	h(\tau, \nu)=\sum_{i=1}^{P} h_{i} \delta\left(\tau-\tau_{i}\right) \delta\left(\nu-\nu_{i}\right),
	\end{equation}
	where $\tau_{i}$, $\nu_{i}$ and $h_{i}$ denote delay, Doppler shift and fade coefficient associated with the $i$th path, respectively. Then, the obtained signal at the receiver can be expressed as
	\begin{equation}
	r(t) = \sum_{i=1}^{P} h_{i}s(t-\tau_{i})e^{j2\pi \nu (t-\tau_{i})} + n(t),
	\end{equation}
	where $n(t)$ denotes Gaussian noise with power spectral density $N_{0}$. 
	
	At the receiver, the dual prototype pulse $g_{rx}(t)$ is used to perform matched filter processing, and then the time-frequency received symbols can be obtained as
	\begin{equation}
	Y\left[n, m\right]=\int r(t) g_{rx}^{*}(t-n T) e^{-j 2 \pi m \Delta f(t-n T)} d t.
	\end{equation}
	Finally, the symbols $Y\left[n, m\right]$ are transformed to the delay-Doppler domain through symplectic finite Fourier transform (SFFT). The input-output relationship of end-to-end system can be formulated as
	\begin{equation}
	y[k,l] = \sum_{k^{\prime}=0}^{N-1} \sum_{l^{\prime}=0}^{M-1} x[k^{\prime},l^{\prime}]h_{\omega}[k-k^{\prime},l-l^{\prime}]+w[k,l],
	\label{eq6}
	\end{equation}
	where $h_{\omega} \in \mathbb{C}^{N \times M}$ denotes the channel impulse response (CIR) in delay-Doppler domain and $w[k,l]$ is a zero-mean Gaussian noise term with variance $N_{0}$. In \cite{8424569}, the explicit formulation of CIR has benn derived. For integer Doppler, there are only $P$ non-zero elements in $h_{\omega}$. On the contrary, fractional Doppler will yield extra inter-Doppler interferences, which will increase the computation complexity of receiver.
	
	\section{Receiver Design}
	\subsection{Canonical SPA Receiver}
	\begin{figure}[!t]
		\centering
		\includegraphics[width=2in]{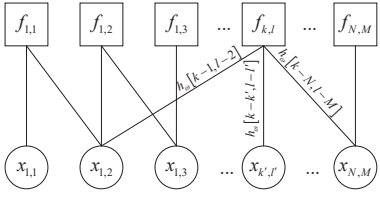}
		\caption{The factor graph of (\ref{eq9}).}
		\label{fig1}
	\end{figure}
	The vectorized form of (\ref{eq6}) can be rewritten as
	\begin{equation}
	\mathbf{y} = \mathbf{H}\mathbf{x} + \mathbf{w},
	\label{eq3-1}
	\end{equation}
	where $\mathbf{y} \in \mathbb{C}^{NM\times 1}$, $\mathbf{H} \in \mathbb{C}^{NM\times NM}$, $\mathbf{x} \in \mathbb{\mathcal{A}}^{NM\times 1}$, and $\mathbf{w} \in \mathbb{C}^{NM\times 1}$. Assuming that
	the data symbols are equally distributed, the optimal MAP detector can be expressed as
	\begin{equation}
	\widehat{\mathbf{x}}=\underset{\mathbf{x} \in \mathcal{A}^{N M} \times 1}{\arg \max }P(\mathbf{x} \mid \mathbf{y},\mathbf{H})
	\end{equation}
	Implementing ML detector requires exponential complexity in $NM$, i.e., $|\mathcal{A}|^{N M}$, where $|\mathcal{A}|$ is the cardinality of $\mathcal{A}$. In Bayesian inference, SPA is an alternative approach to compute the exact posterior probabilities with low complexity. Assuming transmitted symbols are uniformly distributed, the posteriori probability can be factorized as a product of several local functions
	\begin{equation}
	\begin{aligned}
		P(\mathbf{x} \mid \mathbf{y},\mathbf{H}) \propto  p(\mathbf{y} \mid \mathbf{x},\mathbf{H} ) \propto \prod_{k,l}  p\left(y_{k,l} \mid \mathbf{x}, \mathbf{H}^{k,l}\right),
		\label{eq9}
	\end{aligned}
	\end{equation}
	where $\mathbf{H}^{k,l}$ denotes the $(kN+l)$th row vector of $\mathbf{H}$, and
	\begin{equation}
	f_{k,l}(\mathbf{x}) = p\left(y_{k,l} \mid \mathbf{x}, \mathbf{H}^{k,l}\right)
	\propto \exp \left(-\frac{\left|y_{k,l}-\mathbf{H}^{k,l}\mathbf{x}\right|^{2}}{\sigma^{2}}\right).
	\end{equation}
	Portions of the overall factor graph corresponding to (\ref{eq6}) has been given in Fig. \ref{fig1}. Since the graph representing a circular convolution has loops, the application of iterative SPA is required \cite{6789822}. Messages from the factor node $f_{k,l}$ to variable node $x_{k^{\prime},l^{\prime}}$ are straightforward given by
	\begin{equation}
	\begin{aligned}
	\mu^{new}_{f_{k,l} \rightarrow x_{k^{\prime},l^{\prime}}}&\left(x_{k^{\prime},l^{\prime}}\right)= \\ &\sum_{\sim\{x_{k^{\prime},l^{\prime}}\}}\left[f_{k,l}(\mathbf{x}) \prod_{z \in \mathcal{N}(f_{k,l}) \backslash\{x_k^{\prime},l^{\prime}\}} \mu^{old}_{z \rightarrow f_{k,l}}(z)\right],
	\label{eq11}
	\end{aligned}
	\end{equation}
	where the notation $\mathcal{N}(f_{k,l}) \backslash \{x_k^{\prime}, l^{\prime}\}$ denotes the set of variable nodes connected to $f_{k,l}$ excluding the $x_k^{\prime},l^{\prime}$, and the notation $\sum_{\sim\{x_{k^{\prime},l^{\prime}}\}}$ denotes a sum over all variables of local function $f_{k,l}(\mathbf{x})$ excluding $x_{k^{\prime},l^{\prime}}$. 
	
	Messages from variable node $x_{k^{\prime},l^{\prime}}$ to factor node $f_{k,l}$ are given by
	\begin{equation}
		\mu^{new}_{x_{k^{\prime},l^{\prime}} \rightarrow f_{k,l} }\left(x_{k^{\prime},l^{\prime}}\right) = \prod_{g \in \mathcal{N}(x_{k^{\prime},l^{\prime}}) \backslash \{f_{k,l}\}} \mu^{old}_{f_{k,l} \rightarrow g}(x_{k^{\prime},l^{\prime}}),
	\end{equation}
	where the notation $\mathcal{N}(x_{k^{\prime},l^{\prime}}) \backslash \{f_{k,l}\}$ denotes the set of factor nodes connected to $x_{k^{\prime},l^{\prime}}$ excluding the $f_{k,l}$. 
	
	It can be observed that the computational complexity of SPA primarily comes from the summary operator in (\ref{eq11}). Due to the sparsity of delay-Doppler CIR, the number of effective connections for each factor node is significantly less than $NM$. For integer Doppler, calculating $\mu_{f_{k,l} \rightarrow x_{k^{\prime},l^{\prime}}}$ requires collecting messages from $P-1$ edges, so the summary operation involves $|\mathcal{A}|^{P-1}$ terms, which might be feasible  when the number of paths is small. However, fractional Doppler leads to additional inter-Doppler interferences making it prohibitive to establishing ergodicity of the alphabet.
	
	\subsection{Modified Graph and Low Complexity SPA}
	
	\begin{figure}[!t]
		\centering   
		\subfigure[Factor to variable]
		{
			\begin{minipage}[t]{0.4\linewidth}
				\centering         
				\includegraphics[width=0.8\textwidth]{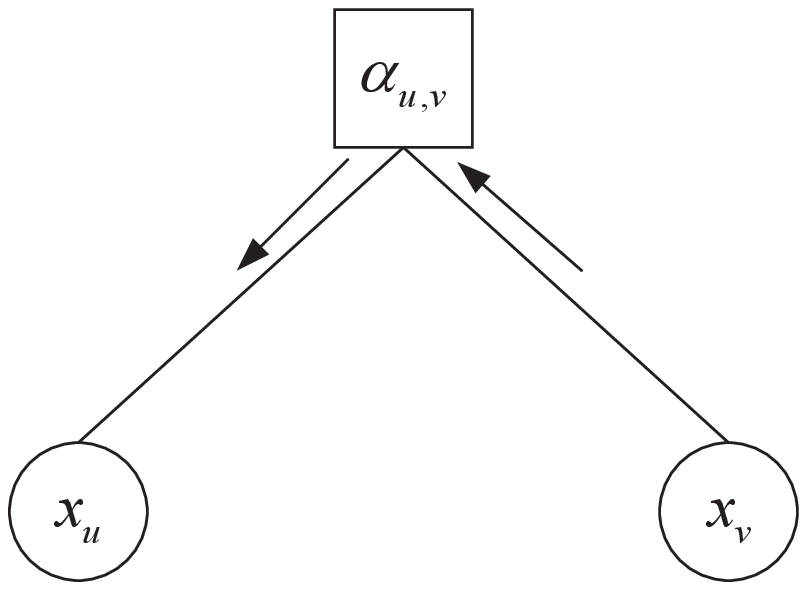}   
			\end{minipage}
		}
		\subfigure[Variable to factor] 
		{
			\begin{minipage}[t]{0.4\linewidth}
				\centering      
				\includegraphics[width=0.8\textwidth]{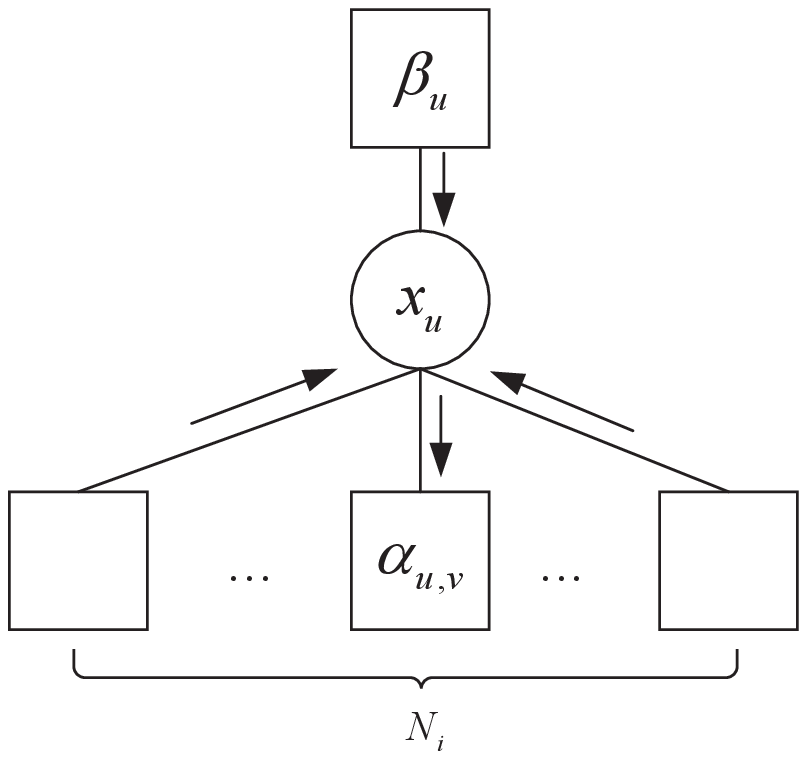}   
			\end{minipage}
		}
		\caption{The factor graph of (\ref{eq14}).} 
		\label{fig2} 
	\end{figure}

	\begin{figure}[!t]
		\centering   
		\subfigure[The location of non-zeros elements in $\mathbf{Q}$.]
		{
			\begin{minipage}[t]{0.4\linewidth}
				\centering         
				\includegraphics[width=1\textwidth]{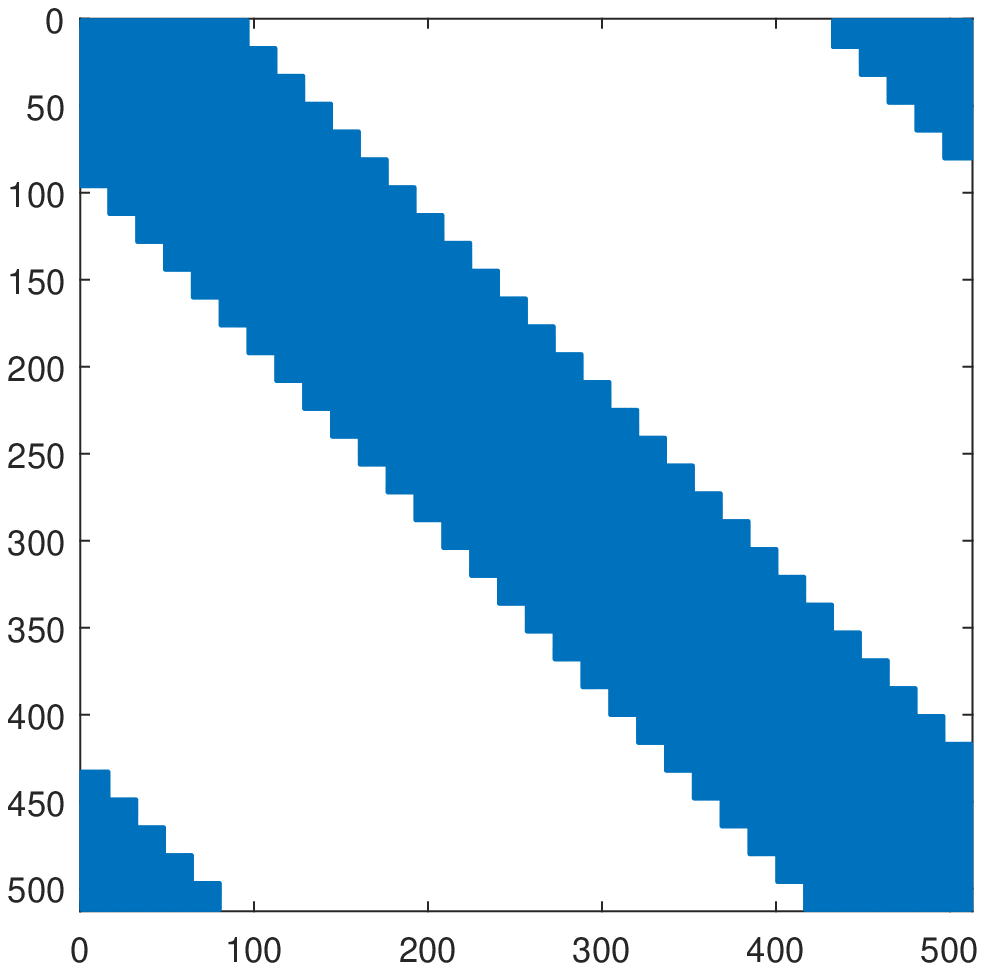}   
			\end{minipage}
		}
		\subfigure[The magnitude of the elements in the first row excluding the  diagonal term.] 
		{
			\begin{minipage}[t]{0.4\linewidth}
				\centering      
				\includegraphics[width=1\textwidth]{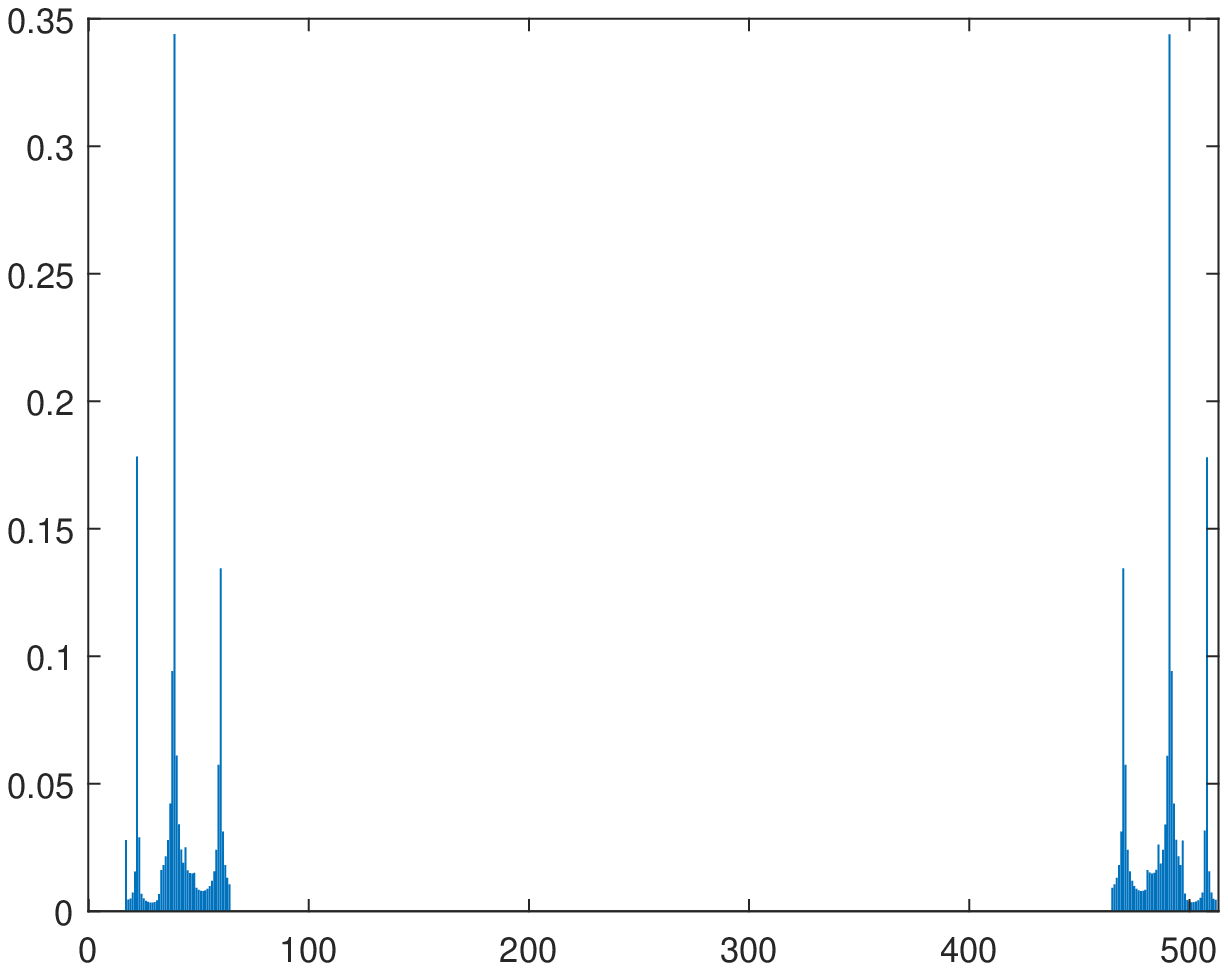}   
			\end{minipage}
		}
		\caption{Structure of matrix $\mathbf{Q}$.} 
		\label{fig3} 
	\end{figure}
		
	To aviod involving overmuch terms in summary operator, we further factorize the posteriori probabilities as 
	\begin{equation}
	\begin{aligned}
		P\left(\mathbf{x} \mid \mathbf{y}, \mathbf{H}\right) & \propto  \exp\left(-\frac{\left(\mathbf{y}-\mathbf{H}\mathbf{x}\right)^{H}\left(\mathbf{y}-\mathbf{H}\mathbf{x}\right)}{\sigma^{2}}\right) \\
		& \propto  \exp \left(-\frac{\left(\mathbf{x}^{H}\mathbf{H}^{H}\mathbf{H}\mathbf{x}\right)-2 \Re\left\{\mathbf{y}^{H}\mathbf{H}\mathbf{x}\right\}}{\sigma^{2}}\right).
		\label{eq13}
	\end{aligned}
	\end{equation}
	Define $\mathbf{Q} = \mathbf{H}^{H}\mathbf{H}$ and $\mathbf{r}=\mathbf{y}^{H}\mathbf{H}$, (\ref{eq13}) can be rewritten as
	\begin{equation}
	\begin{aligned}
		P\left(\mathbf{x} \mid \mathbf{y}, \mathbf{H}\right)  \propto  &\left(\prod_{u\neq v} \exp \left(-x_{u}^{*}Q_{u,v}x_{v}\right)\right) \\
		& \cdot \left(\prod_{u} \exp \left(\frac{-x_{u}^{*}Q_{u,u}x_{u}+2\Re\{r_{u}x_{u}\}}{\sigma^{2}}\right)\right) \\
		& = \prod_{u\neq v} \alpha_{u,v}(x_{u},x_{v}) \prod_{u} \beta_{u}(x_{u})
		\label{eq14}
	\end{aligned}
	\end{equation}
	Here, we use $x_{u}$, $Q_{u,v}$ and $r_{u}$ to denote the elements of $\mathbf{x}$, $\mathbf{Q}$ and $\mathbf{r}$ to simplify formulation. Based on the new factorization, we can obtained the modified graph shown in Fig.\ref{fig2}, and the message computations performed at variable nodes and factor nodes are given by
	\begin{equation}
		\mu^{new}_{\alpha_{u,v} \rightarrow x_{u}}\left(x_{u}\right)= \sum_{x_{v}}\alpha_{u,v}(x_{u},x_{v})\mu^{old}_{x_{v} \rightarrow \alpha_{u,v}}(x_{v}),
		\label{eq15}
	\end{equation}
	\begin{equation}
		\mu^{new}_{x_{v} \rightarrow \alpha_{u,v}}(x_{v}) = \beta_{u}(x_{u})\prod_{g \in \mathcal{N}(x_{u}) \backslash \{\alpha_{u,v}\}} \mu^{old}_{\alpha_{g} \rightarrow x_{u}}\left(x_{u}\right).
	\end{equation}
	Compared with (\ref{eq11}), the improvement of (\ref{eq15}) is that the summary operation only involves one term, which significantly reduces the complexity of the messages from factor nodes to variable nodes, but the more subtle factorization increases the number of factor nodes. It can be seen from Fig.\ref{fig3} that $\mathbf{Q}$ is sparse as well, and since $\mathbf{H}$ is block circulant matrix, each row of $\mathbf{Q}$ is a circulant shift of the first row. Therefore, we can only reserve the strongest $N_{i}$ elements in each row of $\mathbf{Q}$, denoted as $\widetilde{ \mathbf{Q}}$, and the influence of the weak connections will diminish over time. In this way, the cost of calculating overall messages from the factor nodes to the variable nodes is reduced to $N_{i}NM|\mathcal{A}|^2$.
	
	Implementing message propagation in logarithmic domain is an alternative method to avoid multiplication effectively. By taking the logarithm of $\mu^{new}_{\alpha_{u,v} \rightarrow x_{u}}$ and $\mu^{new}_{x_{v} \rightarrow \alpha_{u,v}}$, respectively, we have that
	\begin{equation}
	\begin{aligned}
	\ln\mu^{new}_{\alpha_{u,v} \rightarrow x_{u}}\left(x_{u}\right)= \ln \Biggl\{\sum_{x_{v}} \exp \Big[\ln\alpha_{u,v}(x_{u},x_{v}) \\ 
	+ \ln\mu^{old}_{x_{v} \rightarrow \alpha_{u,v}}(x_{v})\Big]\Biggr\},
	\label{eq17}
	\end{aligned}
	\end{equation}
	\begin{equation}
	\ln \mu^{new}_{x_{v} \rightarrow \alpha_{u,v}}(x_{v}) = \ln \beta_{u}(x_{u})
	+ \sum_{g \in \mathcal{N}(x_{u}) \backslash \{\alpha_{u,v}\}} \ln \mu^{old}_{\alpha_{g} \rightarrow x_{u}}\left(x_{u}\right).
	\end{equation}
	(\ref{eq17}) can be further simplified by using the Jacobian logarithm. We adopt the approximate form $\ln\left(\exp(a)+\exp(b)\right)\approx \max(a,b)$, then it yields the final message propagation scheme
	\begin{equation}
		\bar{\mu}^{new}_{\alpha_{u,v} \rightarrow x_{u}}\left(x_{u}\right) = \max_{x_{v}}\left(-x_{u}^{*}\widetilde{Q}_{u,v}x_{v}+\bar{\mu}^{old}_{x_{v} \rightarrow \alpha_{u,v}}(x_{v})\right),
	\end{equation}
	\begin{equation}
	\begin{aligned}
		\bar{\mu}^{new}_{x_{v} \rightarrow \alpha_{u,v}}(x_{v}) = x_{u}^{*}Q_{u,u}x_{u}&-2\Re\{r_{u}x_{u}\} \\ &+\sum_{g \in \mathcal{N}(x_{u}) \backslash \{\alpha_{u,v}\}}\bar{\mu}^{old}_{\alpha_{g} \rightarrow x_{u}}\left(x_{u}\right).
		\label{eq20}
	\end{aligned}
	\end{equation}
	It can be seen that all multiplications and $\exp$ operations are substituted for additions, which can reduce the computational complexity while avoiding arithmetic overflow. Moreover, our proposed detector does not depend on $\sigma^2$.
	
	In practice, the iterative process cannot always converge, and some approximations we assume will exacerbate the chance of oscillation. One simple way to enhance the convergence is to use damping, i.e., the updated messages is taken to be a weighted average between the old calculation and the new calculation. We set the damped form of messages from the factor nodes to variable nodes as
	\begin{equation}
		\widetilde{\mu}^{new}_{\alpha_{u,v} \rightarrow x_{u}}\left(x_{u}\right) = \lambda\bar{\mu}^{new}_{\alpha_{u,v} \rightarrow x_{u}}\left(x_{u}\right)+(1-\lambda)\widetilde{\mu}^{old}_{\alpha_{u,v} \rightarrow x_{u}}\left(x_{u}\right),
		\label{eq21}
	\end{equation}
	where $\lambda \in [0,1)$ is the damping factor. After $K_{max}$ iterations, the non-normalized marginal probability distribution of each $x_{u}$ is proportional to the addition of all incoming messages at the variable nodes $x_{u}$, which is given by
	\begin{equation}
	\begin{aligned}
		P\left(x_{u}\mid \mathbf{y}, \mathbf{H}\right)  \propto x_{u}^{*}Q_{u,u}x_{u}&-2\Re\{r_{u}x_{u}\} \\ &+\sum_{g \in \mathcal{N}(x_{u})}\bar{\mu}^{K_{max}}_{\alpha_{g} \rightarrow x_{u}}\left(x_{u}\right).
		\label{eq22}
	\end{aligned}
	\end{equation}
	To summarize, the proposed low-complexity SPA is presented in Algorithm 1.
	\begin{figure}[t]
		\removelatexerror
		\begin{algorithm}[H]
			\caption{Low Complexity Sum-Product Algorithm}
			\LinesNumbered
			\KwIn{$N$,$M$,$\mathbf{y}$,$\mathbf{H}$,$N_{i}$,$\mathcal{A}$,$\lambda$ and $K_{max}$.}
			\KwOut{$\widehat{\mathbf{x}}$.}
			Calculate $\widetilde{ \mathbf{Q}}$ and $\mathbf{r}$. Denote $\mathcal{G}_{u}$ as the sets of non-zero positions in the $u$th row of $\widetilde{\mathbf{Q}}$.\\
			Initialize all messages to 0.\\
			\For{i=1:$K_{max}$}{
				\For{u=1:$N\times M$}{
					\For{v in $\mathcal{G}_{u}$}{
						 Update the messages $\widetilde{\mu}^{i}_{\alpha_{u,v} \rightarrow x_{u}}\left(x_{u}\right)$ and $\widetilde{\mu}^{i}_{\alpha_{u,v} \rightarrow x_{v}}\left(x_{v}\right)$ based on (\ref{eq21}).
					}
				}	
			\For{u=1:$N\times M$}{
				Update the messages $\bar{\mu}^{i}_{x_{v} \rightarrow \alpha_{u,v}}(x_{v})$ based on (\ref{eq20}).
			}
			\For{u=1:$N\times M$}{
				 Compute the non-normalized $P\left(x_{u}\mid \mathbf{y}, \mathbf{H}\right)$ based on (\ref{eq22}).
			}
			}
		\end{algorithm}
	\end{figure}

	\section{Simulation Results}
	\begin{figure}[!t]
		\centering
		\includegraphics[width=3in]{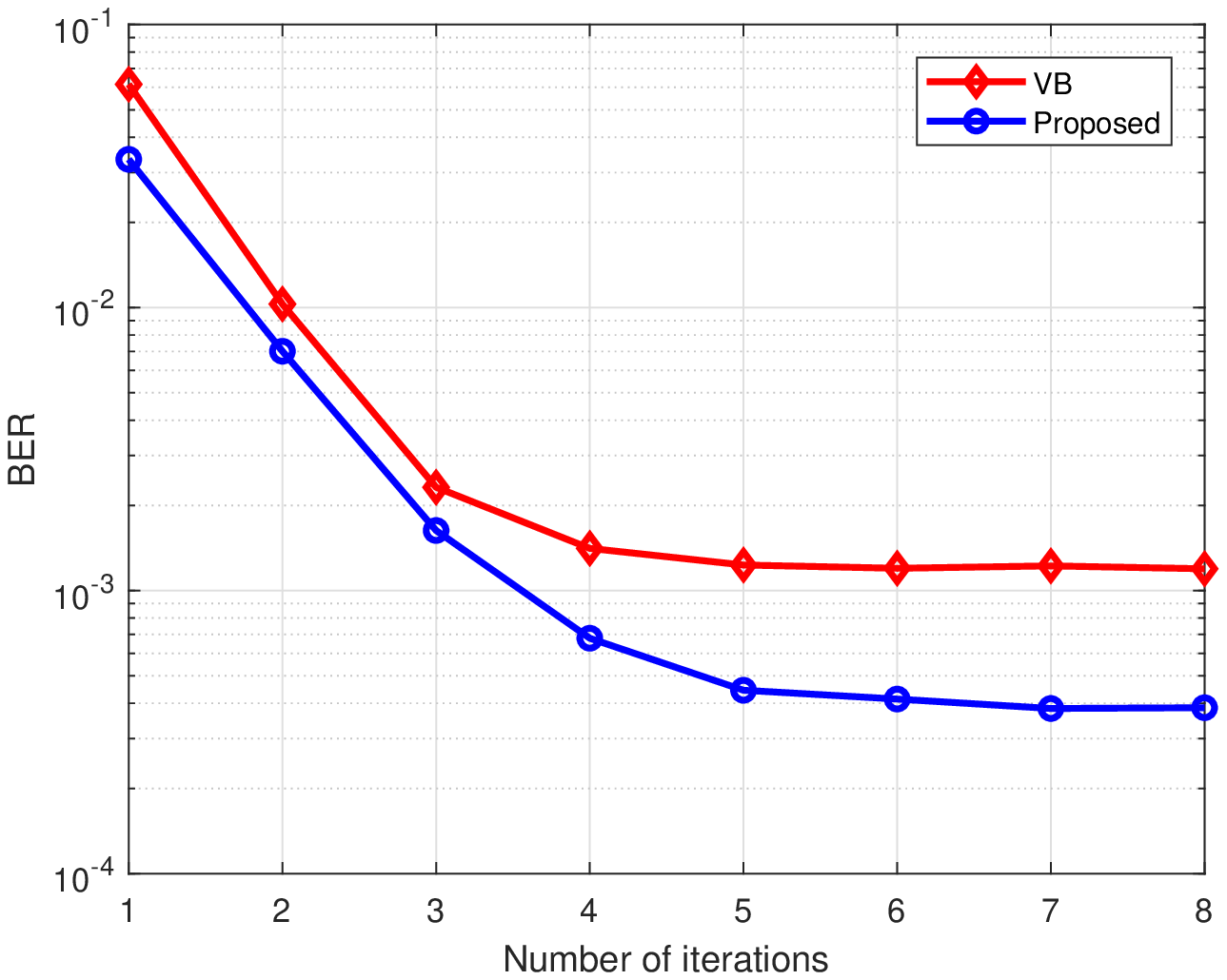}
		\caption{BER performance versus the number of iterations.}
		\label{fig4}
	\end{figure}
	\begin{figure}[!t]
		\centering
		\includegraphics[width=3in]{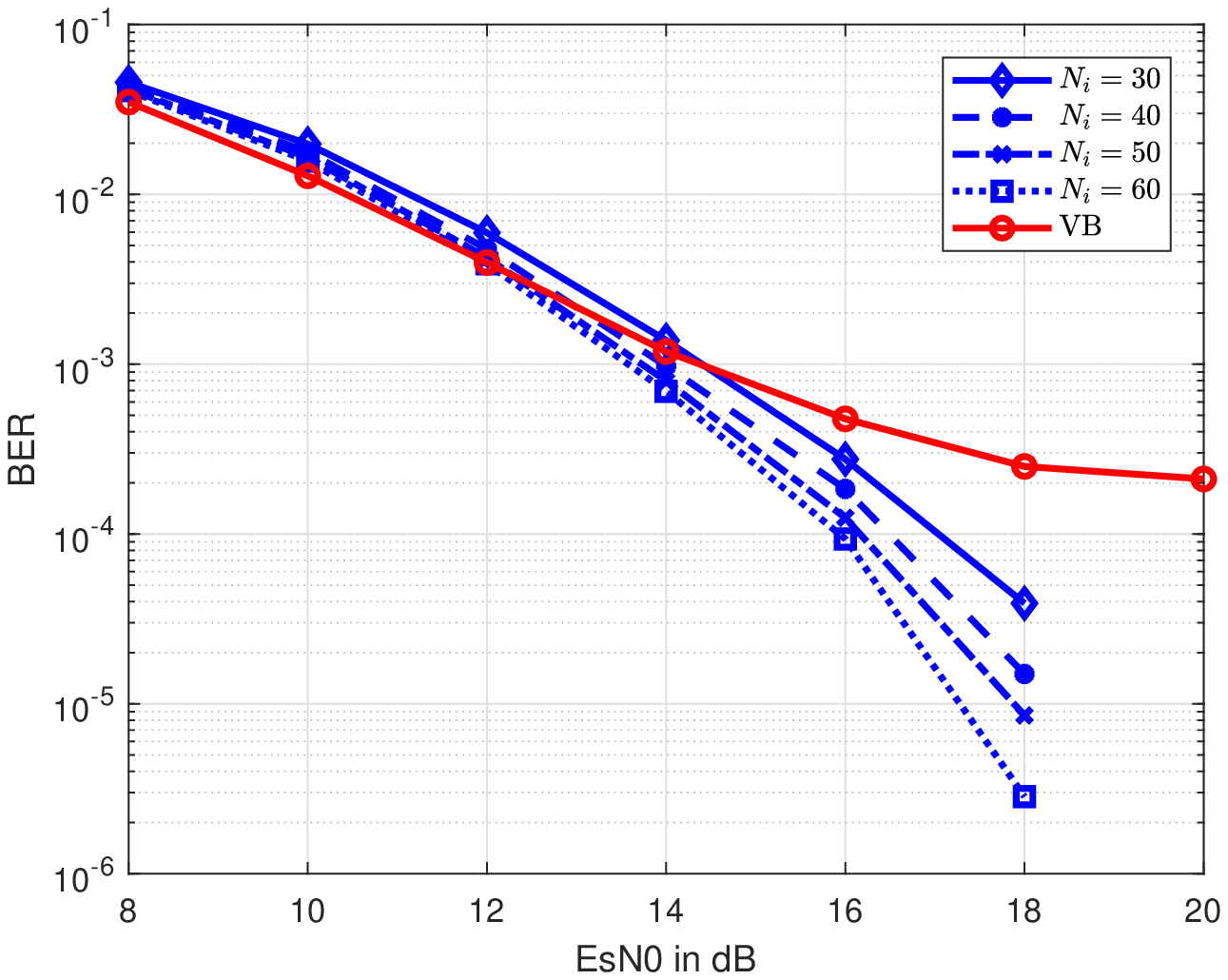}
		\caption{Impact of $N_{i}$ on the BER performance.}
		\label{fig5}
	\end{figure}

	In this section, we illustrate the performance of our proposed algorithm for uncoded OTFS modulation. In our simulation, carrier frequency is 4 GHz and subcarrier separation is 15 kHz. For each OTFS frame, we set $M=128$ and $N=64$. Quadrature phase shift keying (QPSK) modulation is used for symbol mapping. We set the maximum delay index to $l_{\tau_{max}}=10$ and the  the
	maximum Doppler index to $k_{\nu_{max}}=8$,  which is corresponding to speed of the mobile users about 500 km/h. The delay index of the $i$th path is selected from $0, 1, \cdots, l_{\tau_{max}}$ with equal probabilities, and the corresponding Doppler index is randomly selected from $[-k_{\nu_{max}},k_{\nu_{max}}]$. We assume $P=4$ and each channel coefficients $h_{i}$ follow the distribution $\mathcal{C} \mathcal{N}(0,1 / P)$. The a damping factor $\lambda$ is set to 0.5.
	
	First we plot the bit error rate (BER) performance versus the number of iterations in Fig.\ref{fig4}, where the SNR is set to 15 dB and $N_{i}=40$. It can be observed that our proposed algorithm has the same convergence speed as VB detector, but it can achieve a better BER performance compared to that of the VB receiver. In addition, we find that the VB detector cannot completely eliminate interferences even in the absence of noise when the channel coefficients $h_{i}$ tends to be identical, and our proposed only requires more iterations.
	
	In Fig.\ref{fig5}, we compare the BER performance corresponding to differnet $N_{i}$ for OTFS modulation. We can observe that increasing $N_{i}$ leads to a better BER performance, especially the BER performance gap between $N_{i}=60$ and $N_{i}=30$ exceeds an order of magnitude at SNR=18 dB, but it is small at low SNRs. Therefore, there exists a trade-off between the detection performance and the complexity. Moreover, it can be observed that the VB detector only slightly outperforms our proposed algorithm when the SNR is less than 12 dB, however, our proposed algorithm can effectively eliminate the error floor at high SNRs.
	
	\section{Conclusions}
	
	This letter proposed a SPA based receiver for the emerging OTFS modulation with low complexity. To aviod the enumeration of all possible combinations of high-dimensional variables, we design a low complexity receiver by using subtly factorized posteriori probabilities. We further apply Jacobian logarithm to simplify the message propagation processing in logarithmic domain and show that all the multiplication and $\exp$ operation are substituted for addition. Simulation results confirmed the superior BER performance of our proposed algorithm at high SNRs.
	\ifCLASSOPTIONcaptionsoff
	\newpage
	\fi
	
	\bibliographystyle{IEEEtran}
	\bibliography{mybib.bib}
\end{document}